\documentclass[multphys,vecphys]{svmult}

\usepackage{makeidx}         
\usepackage{graphicx}        
\usepackage{multicol}        
\usepackage[bottom]{footmisc}

\makeindex             

\begin{document}

\title*{Generation of Emissions By Fast Particles In Stochastic Media}
\author{Gregory D. Fleishman\inst{1,2}} 
\institute{$^1$National Radio Astronomy Observatory,
Charlottesville, VA 22903 \\ $^2$Ioffe Physico-Technical
Institute, St. Petersburg, 194021 Russia \\
}
%
%
\maketitle

\def\gsim{\ \raise 3pt \hbox{$>$} \kern -8.5pt \raise -2pt \hbox{$\sim$}\ }
\def\lsim{\ \raise 3pt \hbox{$<$} \kern -8.5pt \raise -2pt \hbox{$\sim$}\ }

\section{Introduction}

Interaction of charged particles with each other and/or with
external fields results in emission of electromagnetic radiation.
This article considers the emission arising as fast (nonthermal)
particles move through media with random inhomogeneities. The
nature of these inhomogeneities might be rather arbitrary. One of
the simplest examples of  inhomogeneities is a distribution of the
microscopic particles (atoms or molecules) in an amorphous
substance, so the medium is inhomogeneous at microscopic scales
(of the order of the mean distance between particles), perhaps
remaining uniform (on average) at macroscopic scales.

More frequently, however, real objects are inhomogeneous on
macroscopic scales as well. The irregularities might be related to
the interfaces between inhomogeneities, variations of the
elemental composition, temperature, density, electric, and
magnetic field.

Random inhomogeneities of any of these parameters may strongly
affect in various ways the generation of electromagnetic emission.
For example, the presence of the density inhomogeneities implies
that the dielectric permeability tensor is a random function, as
well as the refractive indices of the electromagnetic eigen-modes.
As a result, the eigen-modes of the uniform medium are not the
same as the eigen-modes of the real inhomogeneous medium. The
irregularities of the electric and magnetic fields affect
primarily the motion of the fast particle (although the effect of
the field fluctuations on the dielectric tensor exists as well).

Below we consider one of many emission processes appearing due to
or affecting by small-scale random inhomogeneities, namely, {\it
diffusive synchrotron radiation} arising as fast particles are
scattered by the small-scale random fields. This emission process
is of exceptional importance since current models of many
astrophysical objects (see, e.g.,
\cite{Jaroshek_etal_2005,2Hohda_2004} and references therein)
imply  generation of rather strong small-scale magnetic fields.
The effect of the inhomogeneities on other emission processes is
discussed briefly as well.

\section{Statistical Methods in the Theory of Electromagnetic Emission}

The trajectories of  charged particles and the fields created by
them are random functions as the particles move through a random
medium. Thus, the use of appropriate statistical methods is
required to describe the particle motion and the related fields.

\subsection{Spectral Treatment of the Random Fields}

For a detailed theory of the random  fields we refer to a
monograph \cite{Toptygin_83} and mention here a few important
points only. To be more specific, let us discuss some properties
of the random \emph{magnetic} fields. Assume that total magnetic
field is composed of regular and random components
$\vec{B}(\vec{r},t)=\vec{B}_0(\vec{r},t) +
\vec{B}_{st}(\vec{r},t)$, such as
$\vec{B}_0(\vec{r},t)=\left<\vec{B}(\vec{r},t)\right>$ and
$\left<\vec{B}_{st}(\vec{r},t)\right>=0$, where the brackets
denote the statistical averaging. Note, that the method of
averaging depends on the problem considered.

The statistical properties of the random field might be described
with a (infinite) sequence of the multi-point correlation
functions, the most important of which is the (two-point)
second-order correlation function
\begin{equation}
\label{Corr_2} K_{\alpha \beta}^{(2)}(\vec{R},T,\vec{r},\tau) =
\left<B_{st,
\alpha}(\vec{r}_1,t_1)B_{st,\beta}(\vec{r}_2,t_2)\right>,
\end{equation}
where $\vec{R} =(\vec{r}_1 + \vec{r}_2)/2$, $\vec{r} =\vec{r}_2 -
\vec{r}_1$, $T =(t_1 + t_2)/2$, and $\tau =t_2 - t_1$.

Since the regular and random fields are statistically independent
($\left<B_0 B_{st}\right> =B_0 \left<B_{st}\right> = 0$), each of
them satisfies the Maxwell equations {\cal separately}. In
particular $\vec{\nabla} \cdot \vec{B}_{st} = 0$, so only two of
three vector components of the random field  are independent.

For statistically uniform random field the Fourier transform of
the correlator $K_{\alpha \beta}^{(2)}(\vec{r},\tau)$ over spatial
and temporal variables $\vec{r}$ and $\tau$ gives rise to the
spectral treatment of the random field
\begin{equation}
\label{Corr_2_spectr} K_{\alpha \beta}(\vec{k}, \omega) = \int
{d\vec{r} d \tau \over (2 \pi)^4} e^{i(\omega \tau - \vec{kr})}
K_{\alpha \beta}^{(2)}(\vec{r},\tau).
\end{equation}
For example, for the isotropic turbulence we can easily find
\begin{equation}
\label{Corr_2_isotr} K_{\alpha \beta}(\vec{k},
\omega)=\frac{1}{2}K(\vec{k}) \delta(\omega -
\omega(\vec{k}))\left(\delta_{\alpha \beta} - {k_{\alpha}
k_{\beta} \over k^2}\right),
\end{equation}
which, in particular, satisfies  the Maxwell equation
$\vec{\nabla} \cdot \vec{B}_{st} = 0$, since the tensor structure
of the correlator is orthogonal to the $\bf k$ vector:
$k_{\alpha}K_{\alpha \beta}=0$.

Although the spectral shape of the correlators is not unique and
may substantially vary depending on the situation, we will adopt
for the  purpose of the model a quasi-power-law spectrum of the
random measures:
\begin{equation}
\label{power_spectr}
 K({\bf k})= \frac{A_{\nu}}{(k_{min}^2+k^2)^{\nu/2+1}},\
 A_{\nu}= {\Gamma(\nu/2+1) k_{min}^{\nu-1} \left<A^2\right> \over 3
 \pi^{3/2}\Gamma(\nu/2-1/2)},
\end{equation}
where $\nu$  is the spectral index of the turbulence, and the
spectrum  $K(\vec{k})$ is normalized to $d^3k$:
\begin{equation}
\label{spectr_norm} \int_{0}^{k_{max}} K(\vec{k}) d^3k =
\left<A^2\right>, \ \hbox{for} \ k_{min} \ll k_{max}, \ \nu >1,
\end{equation}
where $\left<A^2\right>$ is the mean square of the corresponding
measure of the random field, e.g., $\left<B_{st}^2\right>$,
$\left<E_{st}^2\right>$, $\left<\Delta N^2\right>$ etc.

\subsection{Emission by Particle Moving along a Stochastic Trajectory}

The intensity of the emission of the  eigen-mode $\sigma$
\begin{equation}
\label{E_def_sigma} {\cal E}_{\vec{n}, \omega}^{\sigma}=(2 \pi)^6
{\omega^2 n_{\sigma}(\omega) \over c^3} |(\vec{e}_{\sigma} \cdot
\vec{j}_{\omega, \vec{k}})|^2
\end{equation}
depends on the trajectory of the radiating charged particle since
the Fourier transform $\vec{j}_{\omega, \vec{k}}$ of the
corresponding electric current has the form:
\begin{equation}
\label{current} \vec{j}_{\omega, \vec{k}} =  Q
\int_{-\infty}^{\infty}\vec{v}(t) \exp(i\omega t - i
\vec{k}\vec{r}(t)) {dt \over (2 \pi)^4},
\end{equation}
where $Q$ is the charge of the particle. For the stochastic motion
of the particle, we have to substitute (\ref{current}) into
(\ref{E_def_sigma}) and perform the averaging of the corresponding
expression:
\begin{equation}
\label{E_sigma_1} {\cal E}_{\vec{n}, \omega}^{\sigma}= {Q^2
\omega^2 n_{\sigma}(\omega) \over 4 \pi^2 c^3} Re \int_{-T}^{T} dt
\int_{0}^{\infty} d \tau e^{i \omega \tau}
\left<e^{-i\vec{k}[\vec{r}(t+\tau)- \vec{r}(t)]}
(\vec{e}^*_{\sigma} \cdot \vec{v}(t+\tau)) (\vec{e}_{\sigma} \cdot
\vec{v}(t)) \right>,
\end{equation}
where  $2T$ is the total time at which the emission occurs, and
$\vec{e}_{\sigma}$ is the polarization vector of the  eigen-mode
$\sigma$.

It is convenient to perform the averaging denoted by the brackets
with the use of the distribution function of the particle(s)
$F(\vec{r},\vec{p},t)$ at the time $t$ and the conditional
probability $W(\vec{r},\vec{p},t;\vec{r}',\vec{p}',\tau)$ for the
particle to transit from the state $(\vec{r},\vec{p})$ to the
state $(\vec{r}',\vec{p}')$ during the time $\tau$. For
statistically uniform random field we obtain:

\begin{equation}
\label{E_sigma_2} {\cal E}_{\vec{n}, \omega}^{\sigma}= {Q^2
\omega^2 n_{\sigma}(\omega) \over 4 \pi^2 c^3} Re \int_{-T}^{T} dt
\int_{0}^{\infty} d \tau e^{i \omega \tau} \int
d\vec{r}d\vec{p}d\vec{p}' (\vec{e}^*_{\sigma} \cdot \vec{v}')
(\vec{e}_{\sigma} \cdot \vec{v}) F(\vec{r},\vec{p},t)
W_{\vec{k}}(\vec{p},t;\vec{p}',\tau).
\end{equation}
Then, the integration over $\tau$ gives rise to the temporal
Fourier transform of $W$, so the spectrum of emitted
electromagnetic waves is expressed via the spatial and temporal
Fourier transform of the distribution function of the particle in
the presence of the random field.

\subsection{Kinetic Equation in the Presence of Random Fields}

The conditional probability $W$, which substitutes the particle
trajectory in the presence of random fields, can be obtained from
the kinetic Boltzman equation:
\begin{equation}
\label{Eq_Boltzman_L} {\partial f \over \partial t} +
\vec{v}{\partial f \over \partial \vec{r}} + \vec{F}_L{\partial f
\over
\partial \vec{p}} = 0,
\end{equation}
where  $\vec{F}_L=Q \vec{E} + \frac{Q}{c}[\vec{v},\vec{B}]$ is the
Lorentz force, while the electric ($\vec{E}$) and magnetic
($\vec{B}$) fields contain in the general case both regular and
random components. Let us express the Lorentz force as a sum of
these two components explicitly:
\begin{equation}
\label{Sum_force} \vec{F}_L=\vec{F}_R + \vec{F}_{st}.
\end{equation}
Accordingly, we'll seek a distribution function in the form of the
sum of the averaged ($W$) and fluctuating ($\delta W$) components:
\begin{equation}
\label{Sum_dis_fun} f(\vec{r}, \vec{p}, t)=W(\vec{r}, \vec{p}, t)
+ \delta W(\vec{r}, \vec{p}, t).
\end{equation}

Equation for the averaged component $W$ can be derived from
(\ref{Eq_Boltzman_L}) with the Green function method
\cite{Toptygin_83} (magnetic fields only are included below for
simplicity):
\begin{equation}
\label{Eq_W_gen}
{\partial W \over \partial t} + \vec{v}{\partial W \over \partial
\vec{r}} -(\vec{\Omega} \overrightarrow{{\cal O}}) W  =
\frac{1}{2} \left(\frac{Qc}{{\cal
E}}\right)^2\int_{-\infty}^{\infty} d \tau {\cal O}_{\alpha}
T_{\alpha \beta}(\vec{\Delta r}(\tau),\tau) {\cal O}_{\beta}
W(\vec{r} - \vec{\Delta r}(\tau), \vec{p} - \vec{\Delta p}(\tau),
t-\tau),
\end{equation}
where
\begin{equation}
\label{gyro_vec} \vec{\Omega}=\frac{Q\vec{B}c}{\cal E}
\end{equation}
is the vector directed as the magnetic field, whose magnitude
equals  the relativistic gyrofrequency of the charged particle
with energy ${\cal E}$,
\begin{equation}
\label{Corr_T} T_{\alpha \beta}(\vec{r},\tau)=
\left<B_{st,\alpha}(\vec{r}_1,t_1)B_{st,\beta}(\vec{r}_2,t_2)
\right>={\left<B_{st}^2\right> \over 3} \left\{\psi( r)
\delta_{\alpha \beta} + \psi_1( r) { r_{\alpha}  r_{\beta} \over
 r^2} \right\}.
\end{equation}
To derive (\ref{Eq_W_gen}) we transformed the terms with the
magnetic field using the following property of the scalar triple
product:
\begin{equation}
\label{diff_p}
 \frac{Q}{c}[\vec{vB}]\frac{\partial}{\partial \vec{p}}=
 -\frac{Q\vec{B}}{c}[\vec{v}\frac{\partial}{\partial \vec{p}}]=
 -\frac{Q\vec{B}c}{\cal E} \vec{{\cal O}}, \qquad
 \vec{{\cal O}}=[\vec{v}\frac{\partial}{\partial \vec{v}}]
 ,
\end{equation}
where $\vec{{\cal O}}$ is the operator of the velocity angular
variation.

Equation  (\ref{Eq_W_gen}) is rather general and can be applied to
the study of both  emission by fast particles and particle
propagation in the plasma \cite{Toptygin_83}. Further
simplifications of equation  (\ref{Eq_W_gen}) can be done by
taking into account some specific properties of the problems
considered. The theory of wave emission involves a fundamental
measure called the {\it coherence length} (or the formation zone)
that refers to that part of the particle path where the elementary
radiation pattern is formed.

The coherence length is much larger than the wavelength for the
case of relativistic particles, e.g., the coherence length for
synchrotron radiation in the presence of the uniform magnetic
field is $l_s=R_L/\gamma= Mc^2/(QB)$, where $R_L$ is the Larmour
radius, $\gamma= {\cal E}/Mc^2$ is the Lorentz-factor of the
particle. Length $l_s$ is by the factor of $\gamma^2$ larger than
the corresponding wave length. The effect of magnetic field
inhomogeneity on the elementary radiation pattern is specified by
the ratio of the spatial scale of the field inhomogeneity and the
coherence length. If the scale of inhomogeneity is much larger
than the coherence length, the effect of the inhomogeneity is
small and can typically be discarded. However, if the magnetic
field changes noticeably at the coherence length, the
inhomogeneity affects the emission strongly, so the spectral and
angular distributions of the intensity and polarization of the
emission can be remarkably different from the case of the uniform
field.

This means, in particular, that in the presence of magnetic
turbulence with a broad distribution over the spatial scales, the
large-scale spatial irregularities should be considered like the
regular field, while the small-scale fluctuations should be
properly taken into account as the random field.  Since the
variation of the particle speed (momentum) over the correlation
length of the small-scale random field is small, then  we can
adopt $ \Delta \vec{p}(\tau)= 0$, $\Delta \vec{r}(\tau)= \vec{v}
\tau$ in the right-hand-side of Eq. (\ref{Eq_W_gen}) . Then, the
kinetic equation takes the form
\begin{equation}
\label{Eq_W_B_st}
{\partial W \over \partial t} + \vec{v}{\partial W \over \partial
\vec{r}} -(\vec{\Omega} \overrightarrow{{\cal O}}) W =
{\left<B_{st}^2\right> \over 6}\left(\frac{Qc}{{\cal E}}\right)^2
{\cal O}^2 \int_{-\infty}^{\infty} d \tau \psi(v \tau)
 W(\vec{r} - \vec{v}\tau, \vec{p}, t-\tau).
\end{equation}

\subsection{Solution of the Kinetic Equation}

Let us outline the solution of the kinetic equation
(\ref{Eq_W_B_st}) for the averaged distribution function $W$.
First of all, the stochastic field may require splitting onto
large-scale ($\widetilde{\vec{B}}$) and small-scale
($\vec{B}_{st}$) components. To see this, consider a purely
sinusoidal spatial wave of the magnetic field with the strength
$B_0$ and the wavelength $\lambda_0=2\pi/k_0$. If the wavelength
$\lambda_0$ is less than the coherence length $l_{s0}$ calculated
for the emission in uniform field $B_0$, $l_{s0} \sim
Mc^2/(QB_0)$, this wave represents the small-scale field, whose
spatial inhomogeneity is highly important for the emission; in the
opposite case, $\lambda_0 \gsim l_{s0}$, it is the large-scale
field.

The splitting is less straightforward when the random field is a
superposition of the random waves with a quasi-continuous
distribution over the spatial scales. Let us consider the effect
provided by a random magnetic field corresponding to a small range
$\Delta k$ in the spectrum (\ref{power_spectr}) on the charged
particle trajectory. The energy of this magnetic field is
\begin{equation}
\label{dB_fraction}
 \delta {\cal E}_{st} \sim K(\vec{k}) k^2 \Delta k
,
\end{equation}
the corresponding nonrelativistic ``gyrofrequency'' is $\delta
\omega_{st} \sim Q \sqrt{\delta {\cal E}_{st}}/(Mc)$. For a truly
random field, when the harmonics with $\vec{k}$ and $\vec{k}+
d\vec{k}$ are essentially uncorrelated, we can arbitrarily select
the value $\Delta k$ to be small enough to satisfy $\delta
\omega_{st} \ll kc$ for any $k$, so all the independent field
components represent the small-scale field. However, in a more
realistic case the Fourier components of the random field with
similar yet distinct $\vec{k}$ are typically correlated, so they
disturb the particle motion coherently and $\Delta k$ in estimate
(\ref{dB_fraction}) cannot be arbitrarily small any longer.
Accordingly, all components of the random field with $k \lsim
\delta \omega_{st}/c$ (where $\delta \omega_{st}$ is calculated
for the smallest allowable $\Delta k$) must be treated as the
large-scale field.

The large-scale field $\widetilde{\vec{B}}$ together with the
regular field $\vec{B}_0$ specifies the vector $\vec{\Omega}$ in
the left-hand-side of equation (\ref{Eq_W_B_st}):
\begin{equation}
\label{vec_Omega} \vec{\Omega} =
\frac{Q(\vec{B}_0+\widetilde{\vec{B}})c}{\cal E}.
\end{equation}
For the analysis of the emission process (and, respectively, for
the solution of the kinetic equation (\ref{Eq_W_B_st})), we treat
the large-scale field (which is the sum of the regular and
large-scale stochastic fields in a general case) as uniform,
($\vec{\Omega} = const$); the actual inhomogeneity might be taken
into account by averaging the final expressions of the emission if
necessary.

Equation  (\ref{Eq_W_B_st}) in the presence of both uniform and
random magnetic fields has been solved in \cite{Topt_Fl_1987} (sf
\cite{Migdal}):
\begin{equation}
\label{W_w} W_{\vec{k}}= \frac{1}{p^2} \delta(p-p_0) \exp \left[
-i \frac{\omega v}{c}\left(1-\frac{\omega_{pe}^2}{2\omega^2}
\right) \tau \right] w(\vec{\theta}_0,\vec{\theta},\tau),
\end{equation}
where
\begin{equation}
\label{w_def} w(\vec{\theta}_0,\vec{\theta},\tau)=\frac{x}{\pi
\sinh z\tau} \exp \left[-x (\theta^2 + \theta_0^2) \coth z\tau +
2x \vec{\theta}\vec{\theta}_0 \sinh^{-1}z\tau-
\frac{(\vec{\theta}-\vec{\theta}_0)[\vec{n \Omega}]}{2q}-
\frac{\Omega_{\bot}^2 \tau}{4q}\right],
\end{equation}
\begin{equation}
\label{theta_0} \vec{\theta}_0= \frac{\vec{v}_0}{v} - \vec{n}
\left(1 - \frac{\theta_0^2}{2}\right), \qquad \vec{\theta}=
\frac{\vec{v}}{v} - \vec{n} \left(1 - \frac{\theta^2}{2}\right),
\end{equation}
\begin{equation}
\label{x_z_def} x=(1-i)\left(\frac{\omega}{16q}\right)^{1/2};
\qquad z=(1-i)(\omega q)^{1/2},
\end{equation}
\begin{equation}
\label{q_delta_main}
 q(\omega,\theta) = \pi \left(\frac{Qc}{{\cal E}}\right)^2
  \int d\vec{k}'K(\vec{k}') \delta(\omega- (\vec{k}-\vec{k}')\vec{
  v}).
\end{equation}
If there is only random field and no regular field, function $w$
reads
\begin{equation}
\label{w_rand} w(\vec{\theta}_0,\vec{\theta},\tau)= \frac{x}{\pi
\sinh z \tau} \exp \bigg{\{}-x(\theta^2+\theta^2_0) \coth z \tau
+2x \vec{\theta}\vec{\theta}_0 \sinh^{-1}z \tau\bigg{\}},
\end{equation}
in the opposite case where there is no random field, we have
\begin{equation}
\label{w_reg_fin} w(\vec{\theta}_0,\vec{\theta},\tau)=
\delta(\vec{\theta}-\vec{\theta}_0+ [\vec{n}\vec{\Omega}] \tau)
\exp \bigg{\{}\frac{i \omega \tau}{2}\left[\theta_0^2-
\vec{\theta}_0 [\vec{n}\vec{\Omega}] \tau+\frac{\Omega_{\bot}^2
\tau^2}{3}\right]\bigg{\}}.
\end{equation}
Calculation of the emission with the use of this distribution
function leads evidently to the standard expressions of
synchrotron radiation in the uniform magnetic field.

Finally, the distribution function of the free particle (moving
without any acceleration), which does not produce any emission in
the vacuum or uniform plasma, is
\begin{equation}
\label{w_reg_free} w^0(\vec{\theta}_0,\vec{\theta},\tau)=
\delta(\vec{\theta}-\vec{\theta}_0) \exp \bigg{\{}\frac{i \omega
\tau}{2}\theta^2\bigg{\}}.
\end{equation}

\section{Emission by Relativistic Particles in the Presence of Random Magnetic Field}

\subsection{General Case}

Let us consider the energy emitted by a single particle
(regardless the polarization) based on general expression
(\ref{E_sigma_2}), i.e., take the sum of (\ref{E_sigma_2}) over
two orthogonal eigen-modes:
\begin{equation}
\label{E_n_gen} {\cal E}_{\vec{n}, \omega}= {Q^2 \omega^2
\sqrt{\varepsilon(\omega)} \over 2 \pi^2 c^3} Re \int_{-T}^{T} dt
\int_{0}^{\infty} d \tau e^{i \omega \tau} \int
d\vec{r}d\vec{p}d\vec{p}' [\vec{n} \vec{v}']\cdot [\vec{n}\vec{v}]
F(\vec{r},\vec{p},t) W_{\vec{k}}(\vec{p},t;\vec{p}',\tau),
\end{equation}
where we neglected the difference between the two refractive
indices in the magnetized plasma and adopted $n_{\sigma}(\omega)
\approx \sqrt{\varepsilon(\omega)}$.

In the presence of statistically uniform and stationary magnetic
field the emitted energy (\ref{E_n_gen})  is proportional to the
time (on average), although the intensity of emission at a given
direction  $\vec{n}$ depends on time since the angle between the
instantaneous particle velocity  $\vec{v}(t)$  and $\vec{n}$
changes with time as described by the dependence of the function
$F(\vec{r},\vec{p},t)$ on time $t$. This kind of the temporal
dependence is not of particular interest, e.g., it represents
periodic pulses provided by the rotation of the particle in the
uniform magnetic field, so it is more convenient to proceed with
time-independent intensity of radiation emitted into the full
solid angle

\begin{equation}
\label{I_gen_w1}
 I_{\omega}  =  \frac{Q^2 \omega^2}{2 \pi^2c} \sqrt{\varepsilon(\omega)} Re
 \int_0^{\infty} d\tau \exp \left[\frac{i\omega \tau}{2
 \gamma^2}\left(1+\frac{\omega_{pe}^2
 \gamma^2}{\omega^2}\right)\right]
 \int d^2\theta d^2 \theta' ({\vec \theta} {\vec \theta'})
 (w({\vec \theta}, {\vec \theta'}, \tau)-w^0({\vec \theta}, {\vec \theta'},
 \tau)),
\end{equation}
where  $w^0({\vec \theta}, {\vec \theta'}, \tau)$ -- the
distribution function of the free particle (\ref{w_reg_free}),
which does not contribute to the electromagnetic emission (since
the Vavilov-Cherenkov condition cannot be fulfilled in the plasma
or vacuum). Then, calculation of  (\ref{I_gen_w1}), described in
detail in \cite{Topt_Fl_1987}  and reformulated here to more
convenient notations, results in
\begin{equation}
\label{I_gen_3}
 I_{\omega}  = \frac{8Q^2 q(\omega)}{3 \pi  c} \gamma^2 \left(1+\frac{\omega_{pe}^2
 \gamma^2}{\omega^2}\right)^{-1} \Phi_1(s,r) +
 \frac{Q^2 \omega}{4 \pi c \gamma^2} \left(1+\frac{\omega_{pe}^2
 \gamma^2}{\omega^2}\right) \Phi_2(s,r)
,
\end{equation}
where $\Phi_1(s,r)$ and $\Phi_2(s,r)$ stand for the integrals:
\begin{equation}
\label{Fi_1_def}
 \Phi_1(s,r)  = 24s^2 Im \int_0^{\infty}dt
 \exp(-2s_0t)\left[\coth t \exp\left(-2rs_0^3(\coth t - \sinh^{-1}t -
 t/2)\right)-\frac{1}{t}\right]
 ,
\end{equation}
\begin{equation}
\label{Fi_2_def}
 \Phi_2(s,r)  = 4rs^2 Re \int_0^{\infty}dt \frac{\cosh t-1}{\sinh t}
  \exp\left(-2s_0t-2rs_0^3(\coth t - \sinh^{-1}t -
 t/2)\right)
 ,
\end{equation}
which depend on the dimensionless parameters $s_0$,  $s$, $r$:
\begin{equation}
\label{s_def}
 s_0=(1 -i)s  =  \frac{1-i}{8\gamma^2}
  \left(\frac{\omega}{q(\omega)}\right)^{1/2}\left(1+\frac{\omega_{pe}^2
 \gamma^2}{\omega^2}\right)
 ,
\end{equation}
\begin{equation}
\label{r_def}
 r =  32 \gamma^6
  \left(\frac{\Omega_{\bot}}{\omega}\right)^2\left(1+\frac{\omega_{pe}^2
 \gamma^2}{\omega^2}\right)^{-3}
.
\end{equation}

The parameter  $s$ depends on the rate of scattering of the
particle by magnetic inhomogeneities $q(\omega)$, which has the
form
\begin{equation}
\label{q_fin_appr}
 q(\omega)=
 \frac{\sqrt{\pi}\Gamma(\nu/2)\omega_{st}^2\omega_0^{\nu-1}}{3\Gamma(\nu/2-1/2) \gamma^2
 (\alpha^2+\omega_0^2)^{\nu/2}} ,
\end{equation}
for power-law distribution (\ref{power_spectr}) of  magnetic
irregularities over scales: $P(k) = 4\pi K(\vec{k})k^2 \propto
k^{-\nu}$ at $k \gg k_{min}=\omega_0/c$, where
$\omega_{st}^2=Q^2\left<B_{st}^2\right>/(Mc)^2$ is the square of
the cyclotron frequency in the random magnetic field, $\alpha =
(a\omega/2)\left(\gamma^{-2} + \omega_{pe}^2/\omega^2\right)$, $a$
is a factor of the order of unity.

Integrals  (\ref{Fi_1_def}, \ref{Fi_2_def}) cannot be expressed
with elementary functions in the general case. However, there are
convenient asymptotic expressions of these integrals. In
particular, if $r \gg 1$  and  $rs^3 \gg 1$ we have
\begin{equation}
\label{Fi_1_2}
 \Phi_1 \approx -8\pi s^2
 , \qquad
 \Phi_2 \approx 2^{1/3}3^{1/6}\Gamma(2/3)r^{1/3},
\end{equation}
while for $s \gg 1$, $r \ll 1$, functions $\Phi_1$ and $\Phi_2$
contain exponentially small terms:
\begin{equation}
\label{Fi_1_3}
 \Phi_1 \approx 1 - 3\pi^{1/2}2^{3/4}r^{1/4}s^2
 \exp\left(-\frac{8\sqrt{2}}{3\sqrt{r}}\right)
  , \qquad
 \Phi_2 \approx \frac{r}{64s^2} +
 2^{3/4} \pi^{1/2} r^{1/4} \exp\left(-\frac{8\sqrt{2}}{3\sqrt{r}}\right).
\end{equation}
Complementary, for $s \ll 1$  and  $rs^3 \ll 1$ we obtain:
\begin{equation}
\label{Fi_1_1}
 \Phi_1 \approx 6s \frac{1-rs^2}{1+r^2s^4}, \qquad
 \Phi_2 \approx \frac{rs}{2}\frac{1+rs^2}{1+r^2s^4}.
\end{equation}

\subsection{Special cases}

The radiation intensity  (\ref{I_gen_3}) depends on many
parameters, allowing many different parameter regimes. It is clear
that in the absence of the random fields we obtain standard
synchrotron radiation in a uniform magnetic field. Let us consider
here a few interesting cases when the presence of small-scale
random field results in a considerable change of the
emission. 

\subsubsection{Weak Random Magnetic Inhomogeneities Superimposed on Regular Magnetic Field}

\begin{figure} [htbp]
\includegraphics[height=2.03in]{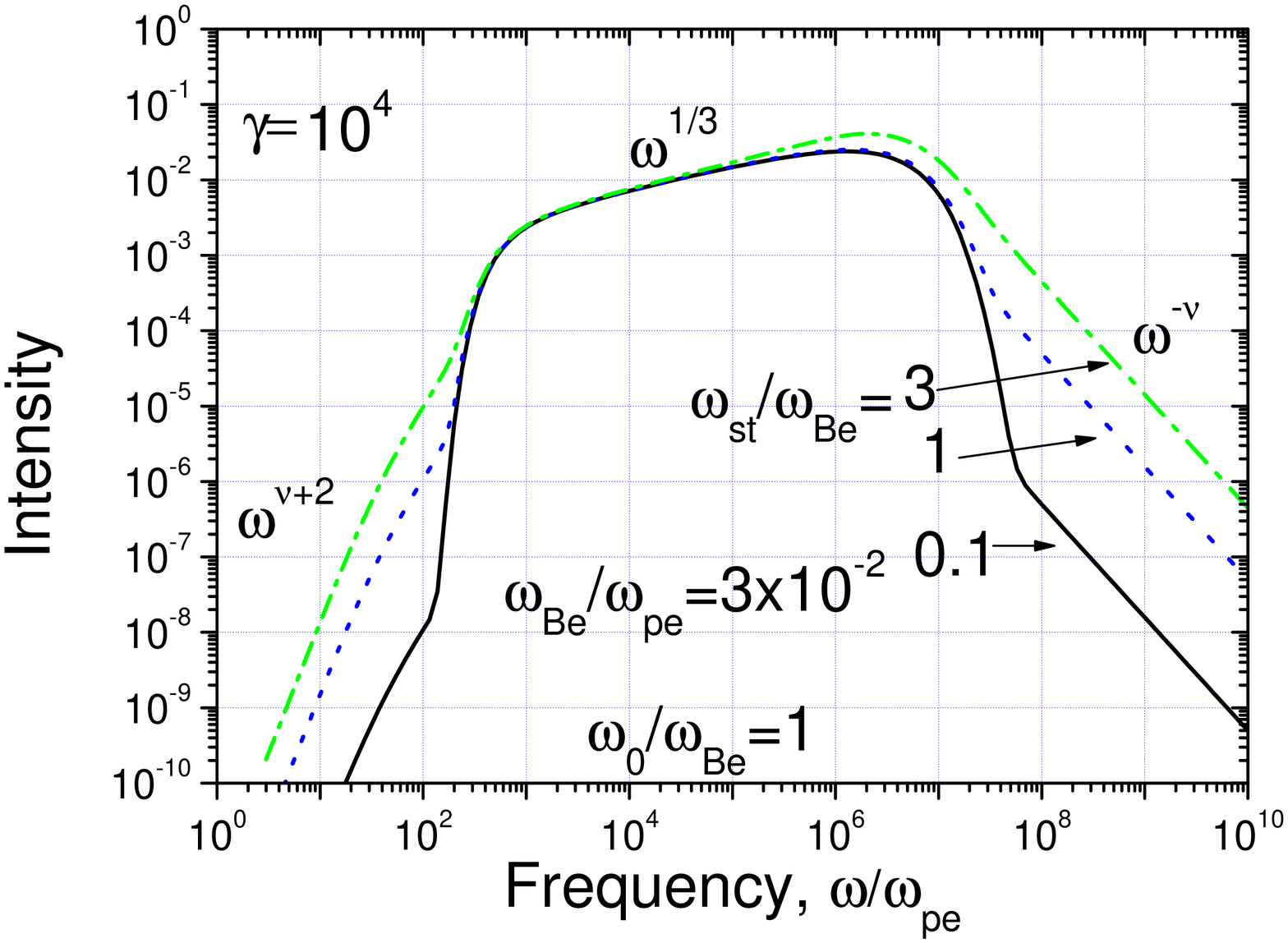}
\includegraphics[height=2.03in]{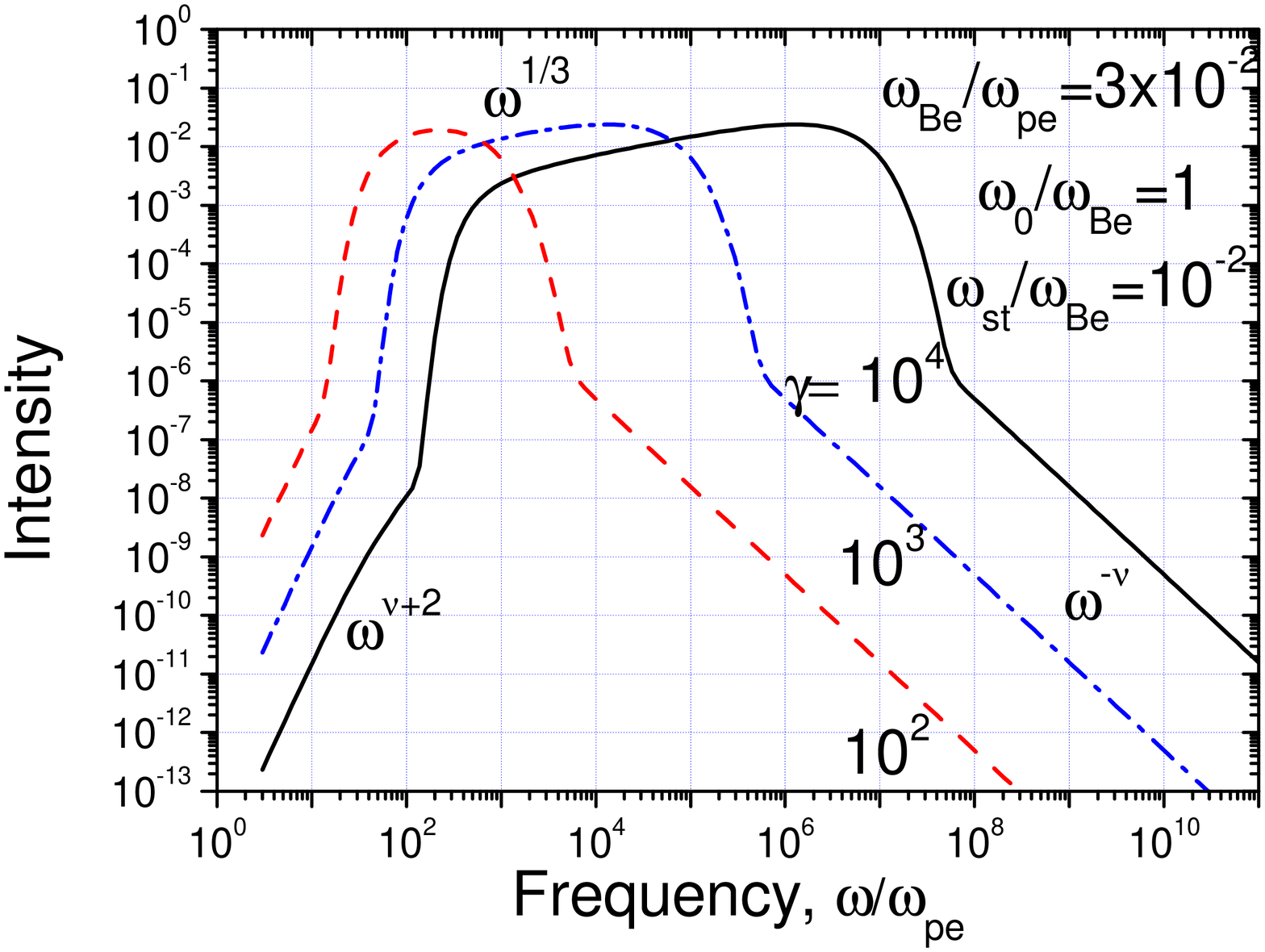}
\caption{Spectra of radiation by a relativistic particle with
$\gamma=10^4$ for differing value of the random magnetic field
(left) and with different $\gamma$ in the presence of weak random
magnetic field $\left<B_{st}^2\right>/B_0^2 = 10^{-4}$ (right).}
\label{fig5_3_2_1}
\end{figure}

Consider the case of weak magnetic irregularities with a broad
(power-law) distribution over spacial scales, $B_{\bot}^2 \gg
\left< \widetilde{B}^2\right>$, so that
\begin{equation}
\label{cond_freq_1}
 \omega_{B} \gg \widetilde{\omega}_{st} \gg \omega_{0},
\end{equation}
here $\widetilde{\omega}_{st}$ is the gyro-frequency related to
the total random field $\left< \widetilde{B}^2\right>^{1/2}$. 

Radiation by highly relativistic particles \cite{Topt_Fl_1987} is
mainly specified by the large-scale (regular plus random) field,
since either $s \gg 1$ or $rs^3 \gg 1$. However, at high
frequencies, where synchrotron radiation decreases exponentially,
the spectrum is controlled by the small-scale field: the spectral
index of radiation is equal to the spectral index of the random
field, see fig. \ref{fig5_3_2_1}.

However, a more interesting regime, which has not been considered
so far, takes place for moderately relativistic (and possibly
non-relativistic) particles moving in a dense plasma (the case
typical for solar and geospace plasmas), when synchrotron
radiation is known to be exponentially suppressed according to
(\ref{Fi_1_3}) by the effect of plasma density (Razin-effect
\cite{Razin,Ginzburg_Syrovatsky_1965}) at all frequencies. The
contribution of the small-scale random field, which we refer to as
{\it diffusive synchrotron radiation}, in this conditions takes
the form:
\begin{equation}
\label{I_case2_full_2}
 I_{\omega}  = \frac{2^{\nu+1}\Gamma(\nu/2)(\nu^2 +7\nu +8)}
 {3 \sqrt{\pi}\Gamma(\nu/2-1/2)(\nu+2)^2(\nu+3)}\ \frac{Q^2}{c} \
 \frac{\omega_{st}^2\omega_{0}^{\nu-1}\gamma^{2\nu}}
 {\omega^{\nu}\left(1+\frac{\omega_{pe}^2 \gamma^2}{\omega^2}\right)^{\nu+1}}.
\end{equation}
\medskip

\begin{figure}[htbp]
\includegraphics[height=2.03in]{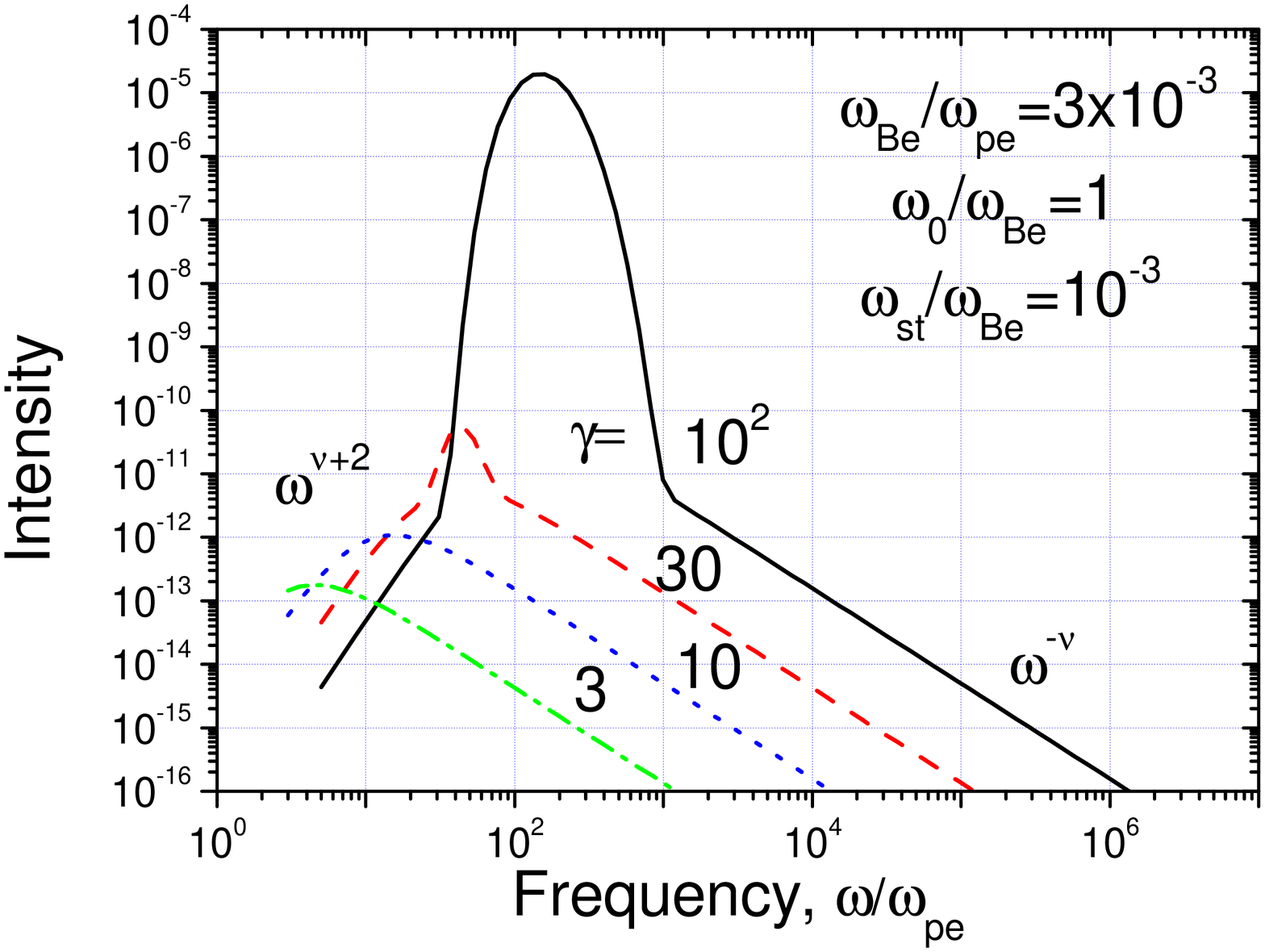}
\includegraphics[height=2.03in]{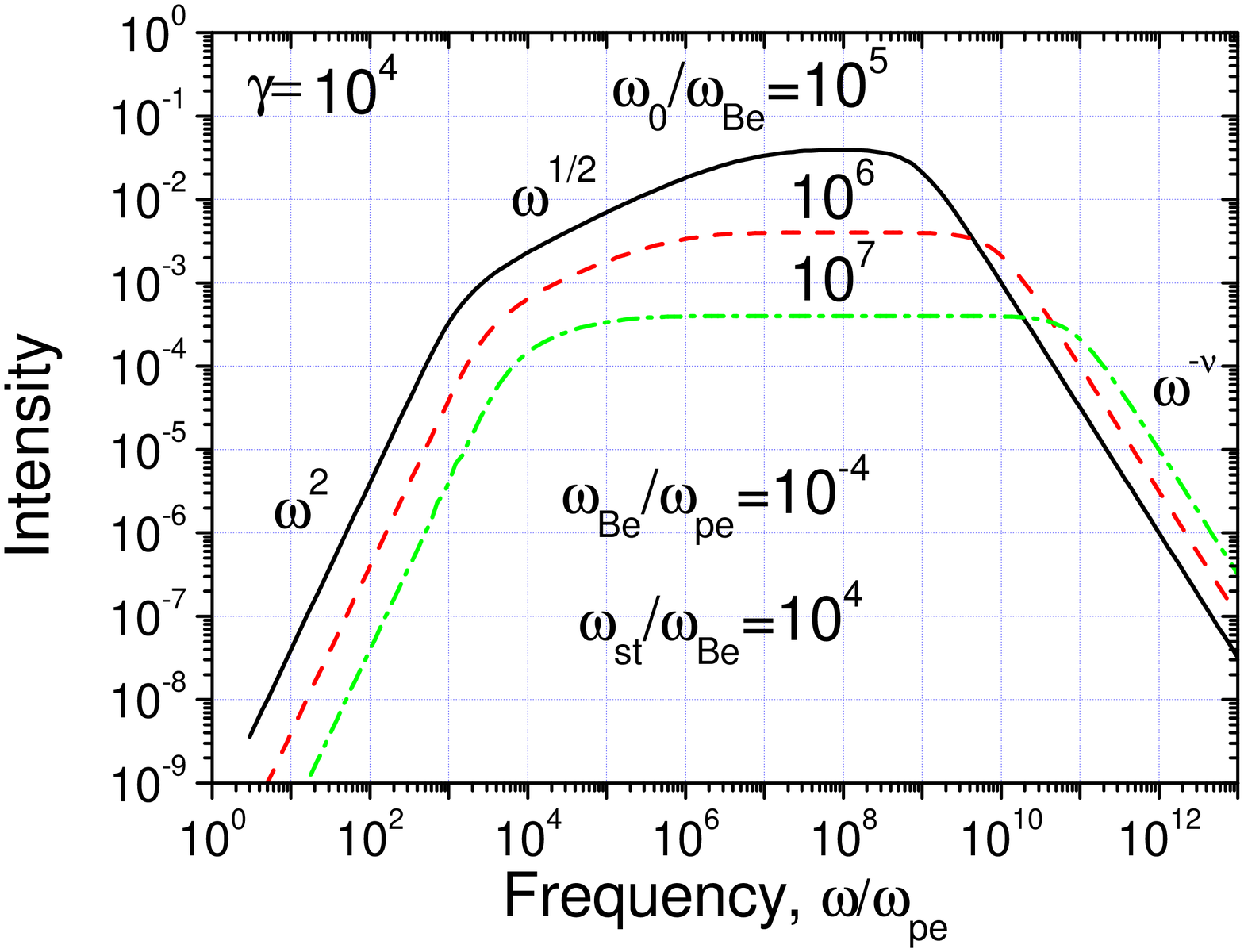}
\caption{Spectra of radiation by a relativistic particle with
different $\gamma$ in a dense plasma in the presence of weak
random magnetic field $\left<B_{st}^2\right>/B_0^2 = 10^{-6}$
(left), and with
 $\gamma=10^4$ in small-scale random magnetic field (right).
If  $\omega_0$ is big enough (e.g., $\omega_0/\omega_{Be}=10^7$ in
the figure) the spectral region provided by multiple scattering,
$\omega^{1/2}$, disappears.} \label{fig5_3_2_3}
\end{figure}

It is important that this radiation decreases with the increase of
the plasma density (plasma frequency) much more slowly (as a
power-law, $\sim \omega_{pe}^{-\nu}$) than synchrotron radiation.
As a result, the diffusive synchrotron radiation can dominate the
entire spectrum even if the random field is much weaker than the
regular field, as is evident from fig. \ref{fig5_3_2_3} left: the
emission by particles with $\gamma \lsim 10$ is defined
exclusively by the small-scale field.

\subsubsection{Small-Scale Magnetic field}

\begin{figure}[htbp]
\includegraphics[height=2.03in]{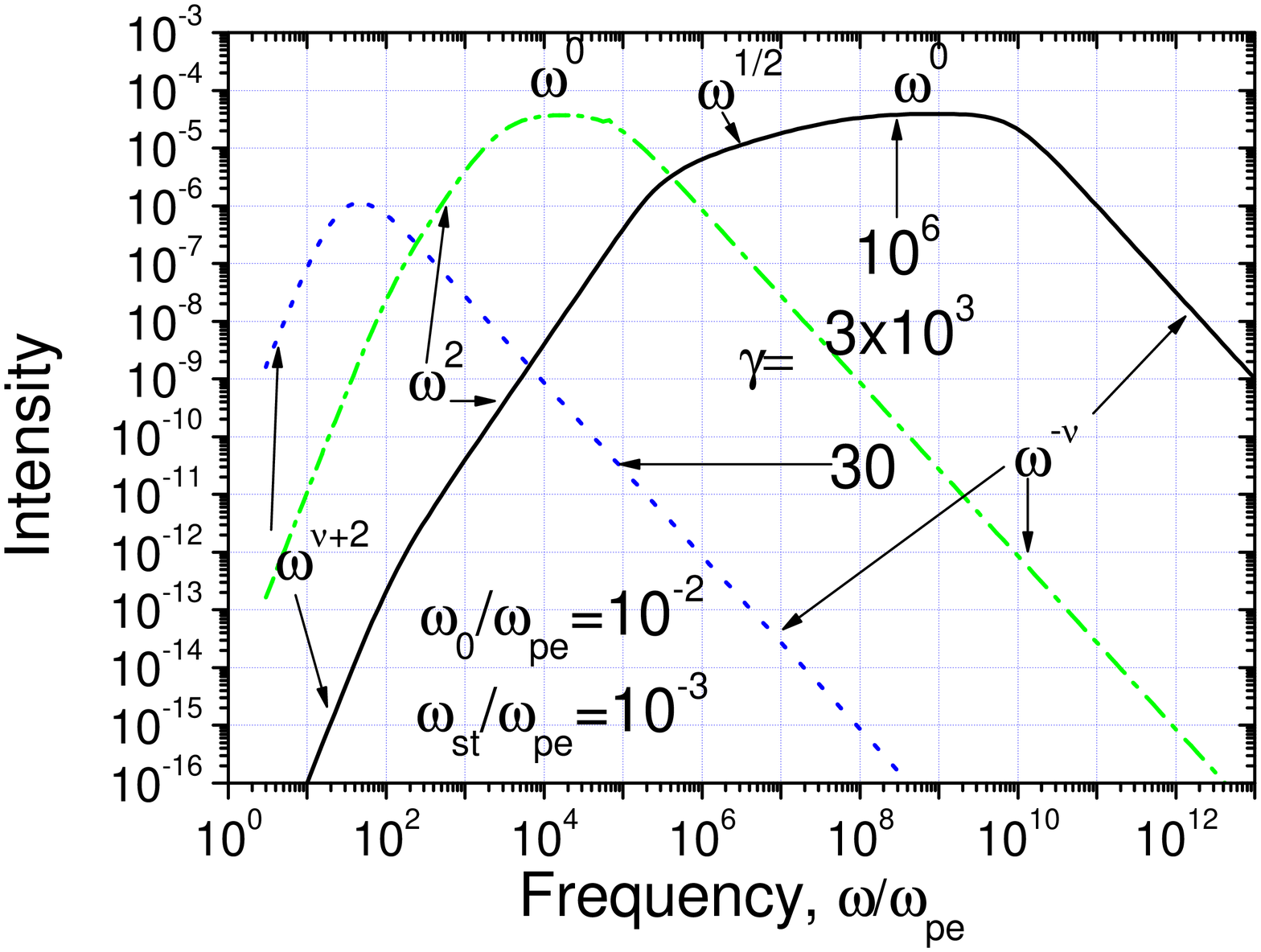}
\includegraphics[height=2.03in]{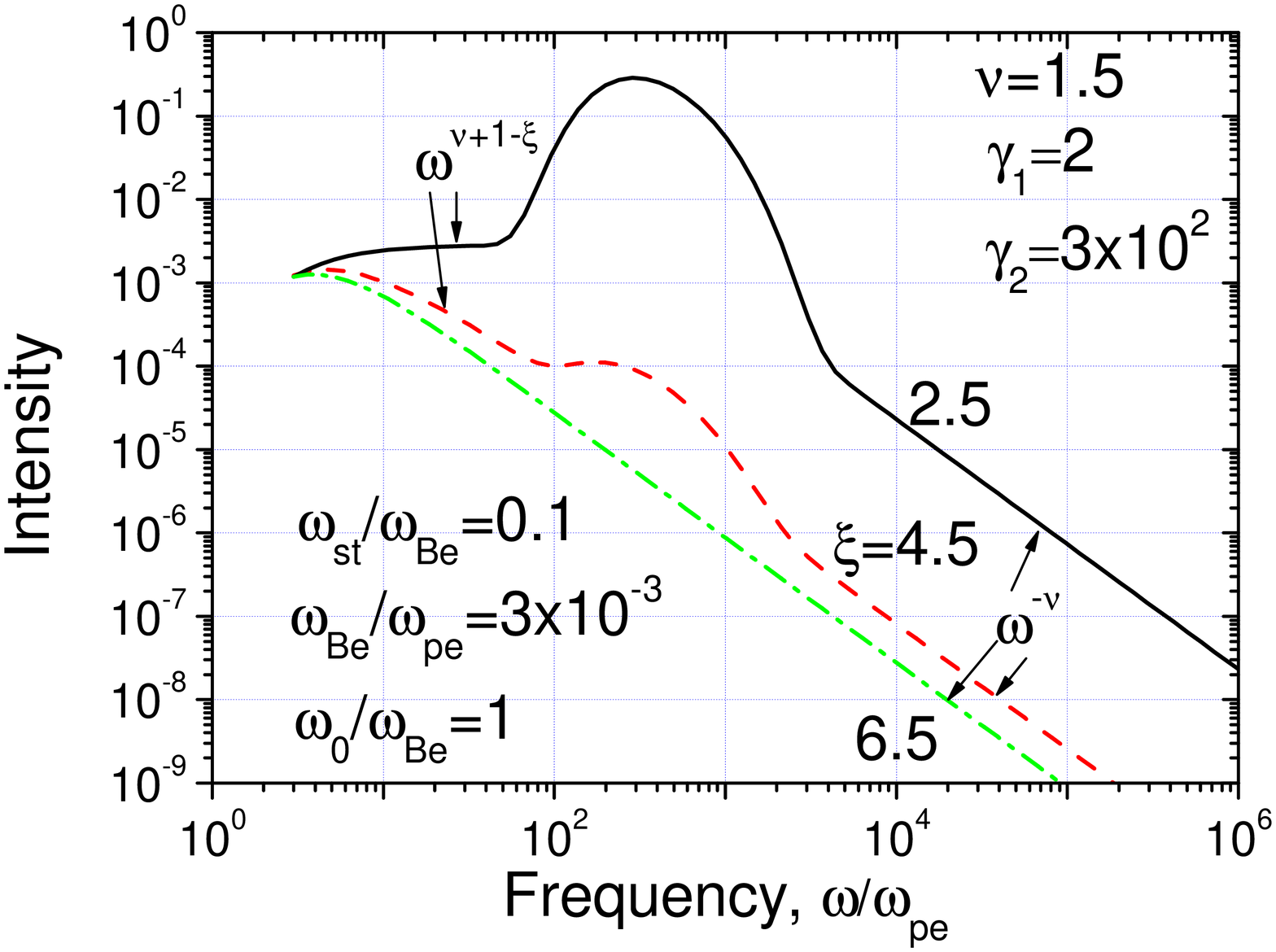}
\caption{\small Spectra of radiation by a relativistic particle
with different $\gamma=30,~3\cdot 10^3,~10^6$ in the presence of
small-scale random magnetic field (left). Emissivity by fast
electron ensemble with different energetic spectra ($\xi=2.5,~4.5,
~6.5$) for the case of dense plasma,
$\omega_{Be}/\omega_{pe}=3\cdot10^{-3}$ (right). }
\label{fig5_3_3_2}
\end{figure}

Consider an extreme case, which is probably relevant for the
physics of cosmological gamma-ray bursts, when there is only
small-scale random magnetic field but no (very weak) regular
field, so that $\omega_{0} \gg \omega_{st}$
\cite{Fleishman_ph_2005}. Now, the parameter $q$ depends
substantially on $\omega_{0}$, the particle motion is similar to
the random walk, so the radiation spectrum is similar to some
extent to bremsstrahlung provided by multiple scattering of the
fast particle by randomly located Coulomb centers. In particular,
the spectrum of diffusive synchrotron radiation can contain a flat
region (as standard bremsstrahlung) and $\propto \omega^{1/2}$
region (like bremsstrahlung suppressed by multiple scattering),
fig. \ref{fig5_3_2_3} right. However, at sufficiently high
frequencies ($\omega > \omega_0 \gamma^2$), the flat spectrum
gives way to a
 power-law region $\propto \omega^{-\nu}$ typical for the diffusive synchrotron
radiation. We should note, that the spectrum depends significantly
on the energy of radiating particle,  fig. \ref{fig5_3_3_2} left,
for low-energy particles some parts of the spectrum (e.g., flat
region) might be missing.

\subsection{Emission by an Ensemble of Particles}

The results presented in the previous section can be directly
applied for mono-energetic electron distributions, which can be
obtained in the laboratory, but are rare exception in  nature
(e.g., astro- and geo- plasmas). Natural particle distributions
can frequently be approximated by a power-law, say, as a function
of dimensionless parameter $\gamma$:
\begin{equation}
\label{distr_fun_power}
  dN_e(\gamma) = (\xi-1)N_e \gamma_1^{\xi-1} \gamma^{-\xi}, \ \ \gamma_1 \le \gamma
  \le \gamma_2,
\end{equation}
where $N_e$  is the number density of relativistic electrons with
energies ${\cal E} \ge mc^2 \gamma_1$, $\xi$ is the power-law
index of the distribution. Evidently, the intensity of incoherent
radiation produced by ensemble (\ref{distr_fun_power}) of
electrons from the unit source volume is
\begin{equation}
\label{power_ans_def}
  P_{\omega} = \int I_{\omega} dN_e(\gamma).
\end{equation}

\subsubsection{Hard Electron Spectrum.}

As we will see, the radiation spectrum produced by an ensemble of
particles differs for hard ($\xi < 2\nu+1$) and soft ($\xi >
2\nu+1$) distributions of fast electrons over energy. Let us
consider first the case of hard spectrum \cite{Topt_Fl_1987},
which is typical, e.g.,  for supernova remnants and radio
galaxies. Assuming the small-scale field to be small compared with
the regular field, we may expect the contribution of diffusive
synchrotron radiation to be noticeable only in those frequency
ranges where synchrotron emission is small.

In particular, at low frequencies $\omega \ll
\max(\omega_{pe}^2/\omega_{B\bot},
\omega_{pe}\sqrt{\omega_{pe}\gamma_1/\omega_{B\bot}})$,
synchrotron radiation is suppressed by the effect of density.
Diffusive synchrotron radiation is produced by relatively low
energy electrons at these frequencies, each electron produces the
emission according to (\ref{I_case2_full_2}), which peaks at
$\omega \sim \omega_{pe}\gamma$. Evaluation of the integral
(\ref{power_ans_def}) gives rise to
\begin{equation}
\label{P_low1_random_ans}
  P_{\omega} \simeq \frac{(\xi-1)\Gamma(\nu/2)(\nu^2 +7\nu +8)}
 {3 \sqrt{\pi}\Gamma(\nu/2-1/2)(\nu+2)^2(\nu+3)}\ \frac{e^2N_e \gamma_1^{\xi-1}}{c} \
 \frac{\omega_{st}^2\omega_{0}^{\nu-1}}
 {\omega_{pe}^{\nu}}\left(\frac{\omega}{\omega_{pe}}\right)^{\nu+1-\xi}
\end{equation}
in agreement with \cite{Nik_Tsyt_1979}. This spectrum can either
increase or decrease with frequency depending on spectral indices
$\nu$ and $\xi$. This expression holds at the frequencies $\omega
\gg \omega_{pe}\gamma_1$. If there are no particles with
$\gamma<\gamma_1$, the spectrum at even lower frequencies drops as
$P_{\omega} \propto \omega^{\nu+2}$:
\begin{equation}
\label{P_low2_random_ans}
 P_{\omega}  = \frac{(\xi-1)2^{\nu+1}\Gamma(\nu/2)(\nu^2 +7\nu +8)}
 {3(\xi+1) \sqrt{\pi}\Gamma(\nu/2-1/2)(\nu+2)^2(\nu+3)}\ \frac{e^2N_e }{c} \
 \frac{\omega_{st}^2\omega_{0}^{\nu-1}\omega^{\nu+2}}{\gamma_1^{2}\omega_{pe}^{2\nu+2}}.
\end{equation}

\begin{figure} [htbp]
\includegraphics[height=2.03in]{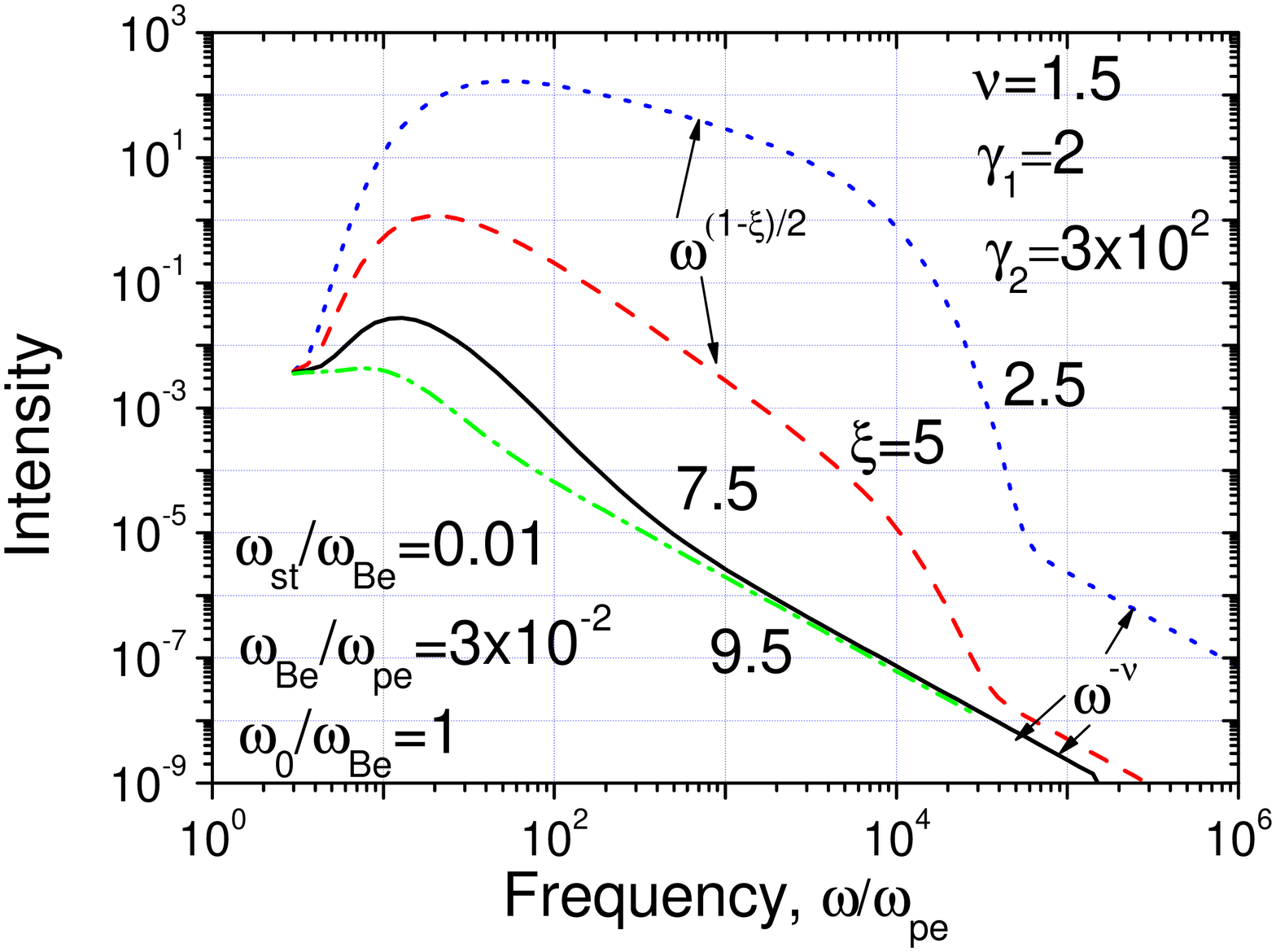}
\includegraphics[height=2.03in]{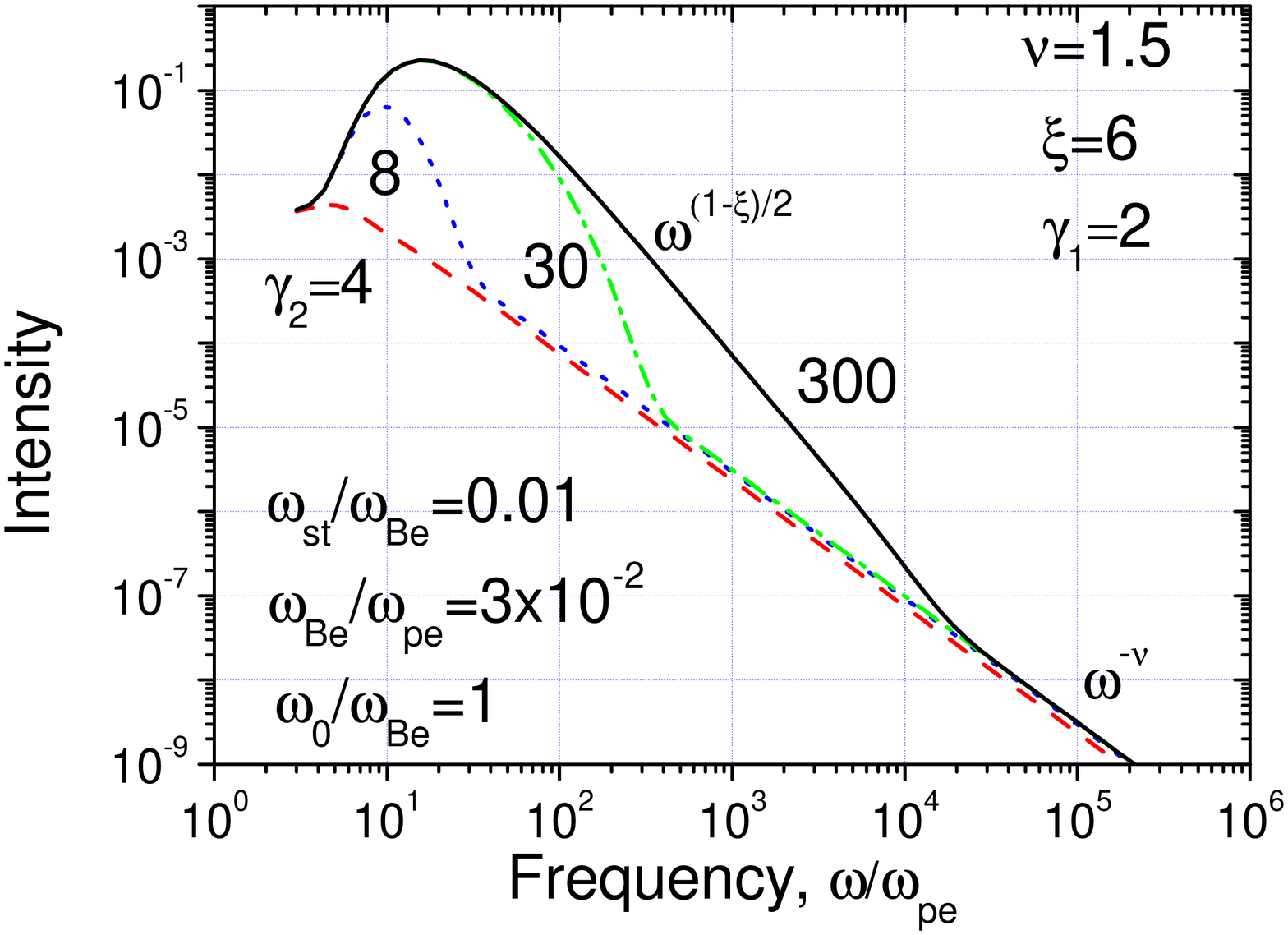}
\caption{\small Left: Same as in fig. \ref{fig5_3_3_2}, right, for
less dense plasma, $\omega_{Be}/\omega_{pe}=3\cdot10^{-2}$. The
contribution from the uniform field (synchrotron radiation)
decreases for softer electron spectra  (i.e., as $\xi$ increases).
Right: Emissivity by fast electron ensemble (with $\xi=6$) from
dense plasma ($\omega_{Be}/\omega_{pe}=3\cdot10^{-2}$) in the
presence of weak magnetic inhomogeneities
$\left<B_{st}^2\right>/B_0^2 = 10^{-4}$ for different high-energy
cut-off values $\gamma_2$. When $\gamma_2$ is small enough, the
uniform magnetic field does not affect the radiation spectrum.}
\label{fig5_4_2_2}
\end{figure}

At high frequencies $\max(\omega_{pe}^2/\omega_{B\bot},
\omega_{pe}\sqrt{\omega_{pe}\gamma_1/\omega_{B\bot}}) \ll \omega
\ll \omega_{B\bot}\gamma_2^2$, where the effect of density is not
important, the spectrum is specified by standard synchrotron
radiation. However, at higher frequencies, $\omega \gg
\omega_{B\bot}\gamma_2^2$ the intensity of synchrotron radiation
decreases exponentially, and the diffusive synchrotron radiation
dominates again. Adding up  contributions from all particles
described by (\ref{I_case2_full_2}) at these frequencies, we
obtain
\begin{equation}
\label{P_high_random_ans}
 P_{\omega}  = \frac{2^{\nu+1}(\xi-1)\Gamma(\nu/2)(\nu^2 +7\nu +8)}
 {3 \sqrt{\pi}(2\nu-\xi+1)\Gamma(\nu/2-1/2)(\nu+2)^2(\nu+3)}\ \frac{e^2N_e \gamma_1^{\xi-1}}{c} \
 \frac{\omega_{st}^2\omega_{0}^{\nu-1}\gamma_2^{2\nu-\xi+1}}{\omega^{\nu}}.
\end{equation}
Thus, power-law spectrum of relativistic electrons with a cut-off
at the energy ${\cal E} = mc^2 \gamma_2$ produce diffusive
synchrotron radiation at high frequencies, whose spectrum shape is
defined by the small-scale field spectrum. Remarkably, the
corresponding flattening in the synchrotron cut-off region has
recently been detected in the optical-UV range for the 3C273 jet
\cite{3C273_jet}, which would imply the presence of relatively
strong small-scale field there in agreement with the model
\cite{2Hohda_2004}.

Although formally spectrum (\ref{P_high_random_ans}) is valid at
arbitrarily high frequencies, there is actually a cut-off related
to the minimal scale of the random field $l_{min}$. Accordingly,
the largest frequency of the diffusive synchrotron radiation is
about $\omega_{max} \sim (c/l_{min})\gamma_2^2$.

\subsubsection{Soft Electron spectrum.}

Let us turn now to the case of sufficiently soft spectra of
electrons, $\xi > 2\nu+1$, which is typical, e.g., for many solar
flares. The contribution of synchrotron radiation is described by
standard expression, $P_{\omega} \propto \omega^{-\alpha}$,
$\alpha=(\xi-1)/2$, which is steeper for the soft spectra than the
spectrum of diffusive synchrotron radiation, $P_{\omega} \propto
\omega^{-\nu}$. Hence, for soft electron spectra, diffusive
synchrotron radiation can dominate even at $\omega <
\omega_{B\bot}\gamma_2^2$. The spectrum of diffusive synchrotron
radiation has the same shape as before but its level is defined by
lower-energy electron contribution:
\begin{equation}
\label{P_random_soft_ans}
 P_{\omega}  = \frac{2^{\nu+1}(\xi-1)\Gamma(\nu/2)(\nu^2 +7\nu +8)}
 {3 \sqrt{\pi}(\xi-2\nu-1)\Gamma(\nu/2-1/2)(\nu+2)^2(\nu+3)}\ \frac{e^2N_e \gamma_1^{2\nu}}{c} \
 \frac{\omega_{st}^2\omega_{0}^{\nu-1}}{\omega^{\nu}}.
\end{equation}
At low frequencies, $\omega \ll \omega_{pe}\gamma_1$, the
radiation is still specified by expression
(\ref{P_low2_random_ans}). One may note that in the case of soft
electron spectra, the emission produced by the electron ensemble
is similar to the emission from a mono-energetic electron
distribution with $\gamma = \gamma_1$, which is the main
difference between the cases of hard and soft electron spectra.

Let us estimate the ratio of the diffusive synchrotron radiation
intensity to the synchrotron  radiation intensity. For simplicity,
we  neglect factors of the order of unity,  assume
$\omega_{0}=\omega_{Be}$ and $\gamma_1 \sim 1$, and introduce
frequency $\omega_*=\omega_{pe}^2/\omega_{Be}$ ($\omega \equiv
(\omega/\omega_*)\omega_*$), around which the synchrotron
emissivity has a peak, then
\begin{equation}
\label{random_synchr_ratio2}
 \frac{P_{diff}}{P_{syn}} \sim
 \frac{\omega_{st}^2}{\omega_{Be}^2}
 \left(\frac{\omega_{pe}}{\omega_{Be}}\right)^{\xi-2\nu-1}
  \left(\frac{\omega}{\omega_{*}}\right)^{\frac{\xi-2\nu-1}{2}}.
\end{equation}
The ratio grows evidently with frequency, so diffusive synchrotron
radiation can become dominant well before the frequency reaches
$\omega_{B\bot}\gamma_2^2$. Moreover, in the case of dense plasma,
$\omega_{pe} \gg \omega_{Be}$, diffusive synchrotron radiation can
dominate at all frequencies under the condition
\begin{equation}
\label{random_synchr_cond}
  \frac{\omega_{st}^2}{\omega_{Be}^2}
 \left(\frac{\omega_{pe}}{\omega_{Be}}\right)^{\xi-2\nu-1}
  > 1,
\end{equation}
even if the random field is small compared with the regular field
$\omega_{st} \ll \omega_{Be}$. On top of this, the radiation
spectrum produced from the dense plasma depends critically on the
highest energy of the accelerated electrons. Indeed, if  $\gamma_2
\ll \omega_{pe}/\omega_{Be}$ (e.g., $\gamma_2=4$ in fig.
\ref{fig5_4_2_2}), then the radiation spectrum is entirely set up
by the small-scale field, in spite of its smallness
($\left<B_{st}^2\right>/B_0^2 = 10^{-4}$ in fig.
\ref{fig5_4_2_2}). Evidently, the standard synchrotron emission
increases and becomes observable as far as $\gamma_2$ increases.

\subsubsection{Diffusive Synchrotron Radiation from Solar Radio Bursts?}

According to microwave and hard X-ray observations of solar
flares, the energetic spectra of accelerated electrons are
frequently rather soft \cite{Nita2004,Kundu_1994}. Consequently,
the diffusive synchrotron radiation can dominate the microwave
emission for dense enough radio sources.
\begin{figure} [htbp]
\includegraphics[height=3.0in]{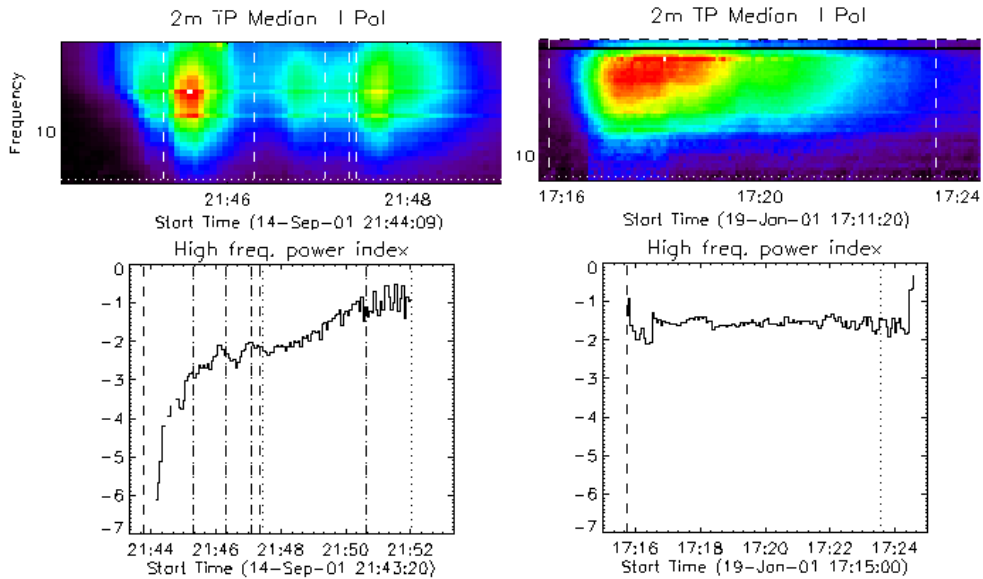}
\caption{\small Two microwave bursts recorded by Owens Valley
Solar Array in the range 1-18 GHz with 40 spectral channels and 4
sec temporal resolution. The first one (left bottom) displays
evident spectral hardening with time, while the second one show
remarkable constancy of the high-frequency spectral slope.
Courtesy by D.E.Gary.} \label{Solar_shake}
\end{figure}
Nevertheless, as a rule the microwave emission from solar flares
meets reasonable quantitative interpretation as synchrotron
(gyrosynchrotron) radiation by moderately relativistic electrons
(moving in non-uniform magnetic field of the coronal loop). An
example of microwave burst produced by gyrosynchrotron emission is
given in fig. \ref{Solar_shake}, left.

Note, that the high-frequency spectral index $\delta$ (the radio
flux is fitted by a power-law, $F \propto f^{\delta}$, at high
frequencies, fig. \ref{Solar_shake}, left bottom) decreases in
value with time. Such spectral evolution typical for solar
microwave bursts is well-understood in the context of the
energy-dependent life time of electrons against the Coulomb
collisions. Indeed, higher energy electrons have longer life time,
which results in spectral hardening of trapped electron
population\cite{Melrose_Brown_1976}, and, respectively, hardening
of the gyrosynchrotron radiation produced
\cite{Melnikov_Magun_1998}.

However, if microwave emission is produced by diffusive
synchrotron radiation as fast electrons interact with small-scale
magnetic (and/or electric) fields, then the radio spectrum is
specified by the spectrum of random fields rather then of fast
electrons. Thus, no spectral evolution (related to electron
distribution modification) is expected. Indeed, there is a
minority of solar microwave bursts, which do not show any spectral
evolution (e.g., no spectral hardening). An example of such a
burst, demonstrating constancy in time of the high-frequency
spectral index, is shown in fig. \ref{Solar_shake}, right.
Curiously, the spectral index is $\delta = -1.5-1.7$ \ in
agreement with standard models \cite{turbulence} and measurements
of the turbulence spectra, e.g., in interplanetary
\cite{Toptygin_83} and interstellar \cite{ISM} space.

Although it has not been firmly proven so far, such microwave
burst are possibly produced by diffusive synchrotron radiation
mechanism. Since (to be  dominant) this mechanism requires
relatively dense plasma at the source site and soft spectra of
accelerated electrons, the observational evidence can be found
from analysis of simultaneous observations of soft and hard  X-ray
emissions from the same flares.

\section{Discussion}

The analysis presented demonstrates the potential importance of
small-scale turbulence in the generation of electromagnetic
emission from natural plasmas. This emission, being reliably
detected and interpreted, provides the most direct  measurements
of small-scale turbulence in the remote sources. The diffusive
synchrotron radiation is only one of the observable effects of the
turbulence on the radio emission. Indeed, the presence of density
inhomogeneities affects the properties of bremsstrahlung, because
the Fourier transform of the square of the electric potential
produced by background charges in a medium depends on spatial
distribution of the charges through the double sum:
\begin{equation}
\label{phi_2_gen}
   \mid  \varphi_{q_0, \vec{q}} \mid^2 \propto
   \left< \sum_{A,B}e^{-
 i\vec{q}(\vec{R}_A-\vec{R}_B)}
 \right>=
 \left[N + (2\pi)^3\frac{|\Delta N|^2_{\vec{q}}}{V}\right],
\end{equation}
where $\vec{R}_A$ and $\vec{R}_B$ are the radius-vectors of
particles $A$ and $B$ respectively, $|\Delta N|^2_{\vec{q}}$ is
the spectrum of the inhomogeneities, $V$ is the volume of the
system. In the uniform amorphous media the positions of various
particles are uncorrelated and this double sum turns to be equal
the total number of particles $N$. However, the macroscopic
inhomogeneities make the positions correlated, so the double sum
deviates from $N$. The second term in (\ref{phi_2_gen}) gives rise
to {\it coherent bremsstrahlung}, which dominates incoherent
bremsstrahlung in a certain spectral range \cite{Pl_Topt_Fl_1990}.

Another important radiation process in the turbulent plasma is
{\it transition radiation} arising as fast particles interact with
small-scale density inhomogeneities of the background plasma (see
\cite{Pl_Fl_2002} and references therein), whose potential
importance for ionospheric conditions has been pointed out long
ago \cite{Erm_Trakht_1981} (see also discussion in
\cite{LaBelle}). This emission process, giving rise to enhanced
low-frequency (at lower frequencies than accompanying synchrotron
emission) continuum radio emission, has recently been reliably
confirmed in a subclass of two-component solar radio bursts
\cite{RTR,RTR_letter}. This finding is of particular importance
for diagnostics of the number density, the level of small-scale
turbulence, and the dynamics of low-energy fast particles in solar
flares.

In addition, the turbulence can also affect the \emph{coherent
emissions} from unstable electron populations \cite{Fl_2004},
e.g., providing strong broadening (or splitting) of the spectral
peaks generated by \emph{electron cyclotron maser} (ECM) emission.
The typical bandwidth of the broadened ECM peaks and its
distributions are found to be quantitatively consistent with those
observed for narrowband solar radio spikes \cite{Fl_2004}.

The National Radio Astronomy Observatory is a facility of the
National Science Foundation operated under cooperative agreement
by Associated Universities, Inc. This work was supported in part
by the Russian Foundation for Basic Research, grants No.
03-02-17218, 04-02-39029. I am strongly grateful to T.S.Bastian
for his numerous comments to the paper.

\end{document}